\documentclass[english,pra,aps,amsmath,amssymb,showpacs,preprint]{revtex4}
\usepackage[T1]{fontenc}
\usepackage[latin9]{inputenc}
\usepackage{amsmath}
\usepackage{amssymb}
\usepackage{graphicx}
\usepackage{esint}

\makeatletter
\@ifundefined{textcolor}{}
{%
 \definecolor{BLACK}{gray}{0}
 \definecolor{WHITE}{gray}{1}
 \definecolor{RED}{rgb}{1,0,0}
 \definecolor{GREEN}{rgb}{0,1,0}
 \definecolor{BLUE}{rgb}{0,0,1}
 \definecolor{CYAN}{cmyk}{1,0,0,0}
 \definecolor{MAGENTA}{cmyk}{0,1,0,0}
 \definecolor{YELLOW}{cmyk}{0,0,1,0}
}

\usepackage{subfigure}\usepackage{hyperref}\usepackage{float}\hypersetup{
unicode=false,
pdftoolbar=true,
pdfmenubar=true,
pdffitwindow=false,
pdfstartview={FitH},
pdftitle={Mytitle},
pdfauthor={Author},
pdfsubject={Subject},
pdfcreator={Creator},
pdfproducer={Producer},
pdfkeywords={keywords},
pdfnewwindow=true,
colorlinks=true,
linkcolor=red,
citecolor=blue,
filecolor=magenta,
urlcolor=cyan
}

\makeatletter\newcommand{\Rmnum}[1]{\expandafter\@slowromancap\romannumeral #1@}\makeatother

\makeatother

\usepackage{babel}

\makeatother

\usepackage{babel}

\makeatother

\usepackage{babel}

\makeatother

\usepackage{babel}

\makeatother

\usepackage{babel}

\makeatother

\usepackage{babel}

\makeatother

\usepackage{babel}

\makeatother

\usepackage{babel}

\makeatother

\usepackage{babel}

\makeatother

\usepackage{babel}

\makeatother

\usepackage{babel}
\begin{document}

\title{Geometric phases for two-mode squeezed state}

\author{Da-Bao Yang}

\email{bobydbcn@163.com}

\affiliation{Department of fundamental physics, School of Science, Tianjin Polytechnic
University, Tianjin 300387, People's Republic of China}

\author{Ji-Xuan Hou}

\affiliation{Department of Physics, Southeast University, Nanjing 211189, People's
Republic of China}

\date{\today}
\begin{abstract}
Although the geometric phase for one-mode squeezed state had been
studied in detail, the counterpart for two-mode squeezed state is
vacant. It is be evaluated explicitly in this paper. Furthermore,
the total phase factor is in an elegent form, which is just identical
to one term of product of two squeezed operators. In addition, when
this system undergoes cyclic evolutions, the corresponding geometric
phase is obtained, which is just the sum of the counterparts of two
isolated one-mode squeezed state. Finally, the relationship between
the cyclic geomtric phase and entanglement of two-mode squeezed state
is established.
\end{abstract}


\maketitle

\section{Introduction}

\label{sec:introduction}

Squeezed light plays an important role in the development of quantum
optics \cite{walls1983squeezed}. It preserves the minimum uncertainty
and exhibits non-classcial nature of light, such as sub-Possionian
statistics which can be observed as photon antibunching effect. It
also has many applications in optical communications and detection
of gravitational radiation. It was be generalized to nonlinear case
by Kwek and Kiang \cite{kwek2003nonlinearsqueezed}. But their studies
were just confined to one mode case. Moreover, two mode squeezed state
was studied by Caves and Schumaker systematically \cite{1caves1985twomode,2schumaker1985twomode}.

Since geometric phase had been discovered by Berry \cite{berry1984quantal}
in the quantum system which underwent adiabatic and unitary evolution,
its research exploded. Subsequently, It was extended to non-Abelian
case by Wilczek and Zee \cite{Wilczek1984appearance}. Its nonadiabatic
and cyclic couterpart \cite{aharonov1987phase,anandon1988nonadiabatic}
was studied by Aharonov and Anandan. Soon, by getting red of the condition
of cyclic evolution, it was generalized to a more general case by
Samuel and Bhandari \cite{samuel1988general}, who depended on Pancharatnam's
earlier study \cite{pancharatnam1956connection}. Subsequently, using
kinematic approach, geometric phase was derived by Samuel and Bhandari
\cite{samuel1988general}.

Moreover, the geometric phases also had other more generalization,
such as off-diagonal ones \cite{manini2000off,mukunda2001bargmann,kult2007nonabelian}
and mixed state couterparts \cite{sjoqvist2000mixed,singh2003geometric,tong2004kinematic}.

In addition, geometric phasese also have many applications, which
range from quantum information and computation science \cite{jones1999geometric,duan2001geometric}
to condensed matter phasics \cite{xiao2010Berry}. These context are
covered by many monographs \cite{shapere1989book,bohm2003book,chru2004book}.

Meantime, the interdiscipline between quantum optics and geometric
phase has also researched. Berry phase for coherent and squeezed states
was researched by Chaturvedi, Sriram and Srinivasan \cite{Chaturvedi1987coherent}.
The nonadiabatic geometric phase for squeezed state was studied by
Liu, Hu and Li \cite{liu1998nonadiabatic}. The geometirc phase for
nonlinear coherent and squeezed state in kinematic approach was disscussed
by Yang et. al. \cite{yang2011nonlinear}. However, the above study
are all confined to one-mode case. As to seek for theoretical progress,
the two-mode case will be researched in this paper. Moreover, the
degree of entanglement between the two-mode are to be evaluated.

This paper is organised as follows. In the next section, the features
of two-modes squeezed states and the kinematic approach to geometric
phase will be reviewed. In Sec. III, the geometric phase for two-mode
squeezed state is to be calculated. From the above outcome, when the
system undergoes cyclic evolution, the corresponding result is also
to be obtained. Moreover, the Von Neumann entropy is going to be calculated.
And its relation with geometric phase will also be established. Finally,
a conclusion is drawn in the last section.

\section{Review of two-mode squeezed states and geometric phases}

\label{sec:reviews}

The Hamiltonian for two-mode of electromagnetic field \cite{schumaker1985twophoton}
takes the form,
\begin{equation}
H_{0}=\Omega(a_{+}^{\dagger}a_{+}+a_{-}^{\dagger}a_{-})+\epsilon(a_{+}^{\dagger}a_{+}-a_{-}^{\dagger}a_{-}),\label{eq:Hamiltonian}
\end{equation}
where $\Omega\pm\epsilon$ are the frequencies for the two-mode and
we take $\hbar=1$ for simplicity. Furthermore, $\Omega$ and $\epsilon$
can be regarded as a carrier frequency and a modulation frequency
respectively. And the electromagnetic field are quantized by the following
commutation relations
\[
[a_{+},a_{-}]=[a_{+}^{\dagger},a_{-}^{\dagger}]=0
\]
\[
[a_{+},a_{+}^{\dagger}]=[a_{-},a_{-}^{\dagger}]=1.
\]

The squeezed operator \cite{schumaker1985twophoton} is generalized
to be
\begin{equation}
S(r,\varphi)\equiv\exp[r(a_{+}a_{-}e^{-2i\varphi}-a_{+}^{\dagger}a_{-}^{\dagger}e^{2i\varphi})],\label{eq:squeezedoperator}
\end{equation}
where the real number $r$ is called the squeeze factor and $\varphi$
is a real phase angle. Moreover, the above operator \eqref{eq:squeezedoperator}
is unitary,
\[
S^{-1}(r,\varphi)=S^{\dagger}(r,\varphi)=S(-r,\varphi).
\]

Hence, the squeezed vacuum state is
\begin{equation}
S(r,\varphi)|0\rangle.\label{eq:initialstate}
\end{equation}
Under the Hamiltonian \eqref{eq:Hamiltonian}, it evolves as
\begin{equation}
\begin{array}{ccc}
e^{-iH_{0}t}S(r,\varphi)|0\rangle & = & e^{-iH_{0}t}S(r,\varphi)e^{iH_{0}t}|0\rangle\\
 & = & S(r,\varphi-\Omega t)|0\rangle
\end{array}\label{eq:evolvedstate}
\end{equation}
which uses the following formulas \cite{schumaker1985twophoton}
\[
\exp[-i\epsilon t(a_{+}^{\dagger}a_{+}-a_{-}^{\dagger}a_{-})]S(r,\varphi)\exp[i\epsilon t(a_{+}^{\dagger}a_{+}-a_{-}^{\dagger}a_{-})]=S(r,\varphi)
\]
and
\[
\exp[-i\theta(a_{+}^{\dagger}a_{+}+a_{-}^{\dagger}a_{-})]S(r,\varphi)\exp[i\theta(a_{+}^{\dagger}a_{+}+a_{-}^{\dagger}a_{-})]=S(r,\varphi-\theta).
\]

The geometric phases $\gamma$ \cite{mukunda1993quantum} for arbitrary
time $t$ takes the form
\begin{equation}
\gamma=\arg\langle\psi(0)|\psi(t)\rangle+\int_{0}^{t}\langle\psi(\tau)|H|\psi(\tau)\rangle d\tau.\label{eq:GeometricPhase}
\end{equation}
It is physical reality ,due to it is invariant under gauge transformation.
And it is can be explained as outcome of parallel transportation in
the framework of fiber bundle, i.e., holonomy. That's the reason that
it deserves a name called Geometric Phase.

\section{Evoluations of the geometric phase factor}

\label{sec:Nonadiabatic}

For convenience, instead of calculating the geometric phase, we evaluate
the geometric phase factor,
\begin{equation}
e^{i\gamma}=\frac{\langle\psi(0)|\psi(t)\rangle}{\parallel\langle\psi(0)|\psi(t)\rangle\parallel}e^{i\delta},\label{eq:GeometricPhaseFactor}
\end{equation}
where
\begin{equation}
\delta=\int_{0}^{t}\langle\psi(\tau)|H|\psi(\tau)\rangle d\tau\label{eq:NegativeDynamicalPhase}
\end{equation}
which is identical to negative the dynamical phase.

At first, let's calculate the inner product
\begin{equation}
\begin{array}{ccc}
\langle\psi(0)|\psi(t)\rangle & = & \langle0|S^{\dagger}(r,\varphi)e^{-iH_{0}t}S(r,\varphi)|0\rangle\\
 & = & \langle0|S^{\dagger}(r,\varphi)S(r,\varphi-\Omega t)|0\rangle,
\end{array}\label{eq:TotalPhaseFactor}
\end{equation}
which ueses Eq. \eqref{eq:evolvedstate}. In order to work out the
total phase, the following formula \cite{schumaker1985twophoton}
is very useful
\begin{equation}
S^{\dagger}(r^{\prime},\varphi^{\prime})S(r^{\prime\prime},\varphi^{\prime\prime})=e^{-i\Theta}S(R,\Phi-\Theta)R(\Theta),\label{eq:ProductDecomposition}
\end{equation}
where the above parameters satisfy the matrix equation
\begin{equation}
C_{R,\Phi}e^{i\Theta\sigma_{3}}=C_{r^{\prime\prime},\varphi^{\prime\prime}}C_{-r^{\prime},\varphi^{\prime}},\label{eq:Parameters}
\end{equation}
where the matrix $C_{r,\varphi}$ is defined by
\begin{equation}
C_{r,\varphi}=\left(\begin{array}{cc}
\cosh r & e^{2i\varphi}\sinh r\\
e^{-2i\varphi}\sinh r & \cosh r
\end{array}\right)\label{eq:Matrix}
\end{equation}
and $\sigma_{3}$ is the famous Pauli matrix in the $z$ direction.
By substituting Eq. \eqref{eq:ProductDecomposition} into Eq. \eqref{eq:TotalPhaseFactor},
one obtains
\[
\langle\psi(0)|\psi(t)\rangle=\langle0|e^{-i\Theta}S(R,\Phi-\Theta)R(\Theta)|0\rangle,
\]
where
\[
R(\Theta)=exp[-i\Theta(a_{+}^{\dagger}a_{+}+a_{-}^{\dagger}a_{-})].
\]
By use of the explicit decomposition of squeezed operator \cite{schumaker1985twophoton}
\begin{equation}
S(R,\Psi)=\left(\cosh R\right)^{-1}e^{-a_{+}^{\dagger}a_{-}^{\dagger}e^{2i\Psi\tanh R}}e^{-(a_{+}^{\dagger}a_{+}+a_{-}^{\dagger}a_{-})\ln(\cosh R)}e^{a_{+}a_{-}e^{-2i\varphi}\tan R},\label{eq:FullDecompostion}
\end{equation}
the total phase can be transformed to be an elegent manner
\begin{equation}
\langle\psi(0)|\psi(t)\rangle=\left(\cosh R\right)^{-1}e^{-i\Theta},\label{eq:TotalPhaseFactorOriginal}
\end{equation}
of which parameters are determinded by Eq. \eqref{eq:Parameters}.
Its explicit form is
\[
\left(\begin{array}{cc}
e^{i\Theta}\cosh R & e^{i(2\Phi-\Theta)\sinh R}\\
e^{i(\Theta-2\Phi)}\sinh R & e^{-i\Theta}\cosh R
\end{array}\right)=\left(\begin{array}{cc}
e^{-i\Omega t(i\cosh2r\sin\Omega t+\cos\Omega t)} & e^{i(2\varphi-\Omega t-\pi/2)}\sinh2r\sin\Omega t\\
e^{-i(2\varphi-\Omega t-\pi/2)}\sinh2r\sin\Omega t & e^{i\Omega t}(\cos\Omega t-i\sin\Omega t\cosh2r)
\end{array}\right).
\]
Therefore, the element $(2,2)$ can tell us the total phase factor
$e^{-i\Theta}$, which take the form
\begin{equation}
\frac{e^{i\Omega t}(\cos\Omega t-i\sin\Omega t\cosh2r)}{(\cos^{2}\Omega t+\sin^{2}\Omega t\cosh^{2}2r)^{1/2}}.\label{eq:FinalTotalPhaseFactor}
\end{equation}

Moreover, let us calculate another term $\delta$ \eqref{eq:NegativeDynamicalPhase}
in the expression of geometric phase \eqref{eq:GeometricPhase}. by
substituting Eq. \eqref{eq:evolvedstate} into Eq. \eqref{eq:NegativeDynamicalPhase},
one can obtain
\[
\delta=\int_{0}^{t}\langle0|S^{\dagger}(r,\varphi-\Omega\tau)HS(r,\varphi-\Omega\tau)|0\rangle d\tau.
\]
By use of the following formulas \cite{schumaker1985twophoton}
\[
\begin{array}{ccc}
S^{\dagger}(r,\eta)a_{+}S(r,\eta) & = & a_{+}\cosh r-a_{-}^{\dagger}e^{2i\eta}\sinh r\\
S^{\dagger}(r,\eta)a_{-}S(r,\eta) & = & a_{-}\cosh r-a_{+}^{\dagger}e^{2i\eta}\sinh r,
\end{array}
\]
the formula for $\delta$ can be simplified as
\begin{equation}
\delta=2\Omega t\sinh^{2}r.\label{eq:FinalNegativeDynamicalPhase}
\end{equation}

Finally, by inserting Eq. \eqref{eq:FinalTotalPhaseFactor} and \eqref{eq:FinalNegativeDynamicalPhase}
into Eq. \eqref{eq:GeometricPhaseFactor}, the geometric phase is
achieved as
\[
e^{i\gamma}=\frac{^{e^{i\Omega t\cosh2r}(\cos\Omega t-i\sin\Omega t\cosh2r)}}{(\cos^{2}\Omega t+\sin^{2}\Omega t\cosh^{2}2r)^{1/2}}
\]

Now, the cyclic geometric phase will be discussed. From the total
phase factor \eqref{eq:TotalPhaseFactor}, it is not hard to see that
when $\Omega t=2\pi$, the state \eqref{eq:evolvedstate} will undergo
a genuine cyclic evolution of which the final state is exactly the
initial state. In another word, the total phase can be regarded as
zero. hence the geometric phase can by explicitly expressed
\begin{equation}
\gamma_{c}=4\pi\sinh^{2}r\quad mod\quad2\pi,\label{eq:CyclicGeometricPhase}
\end{equation}
which is exactly negative the dynamical phase. Because total phase
vanish and the geometric phase is equal to the difference between
the total phase and dynamical phase. In addtion, from Ref. \cite{yang2011nonlinear},
the geometric phase for isolated one-mode squeezed state is
\[
\gamma_{ic}=2\pi\sinh^{2}r\quad mod\quad2\pi,
\]
where the subscripe index $i$ denotes for mode. So if we confine
cyclic geometric phase to a simple form, combining with Eq. \eqref{eq:CyclicGeometricPhase},
$\gamma_{c}=\gamma_{1c}+\gamma_{2c}$, which reveals the addition
relationship between the two-mode system and the isolated one mode
system.

\begin{figure}
\includegraphics{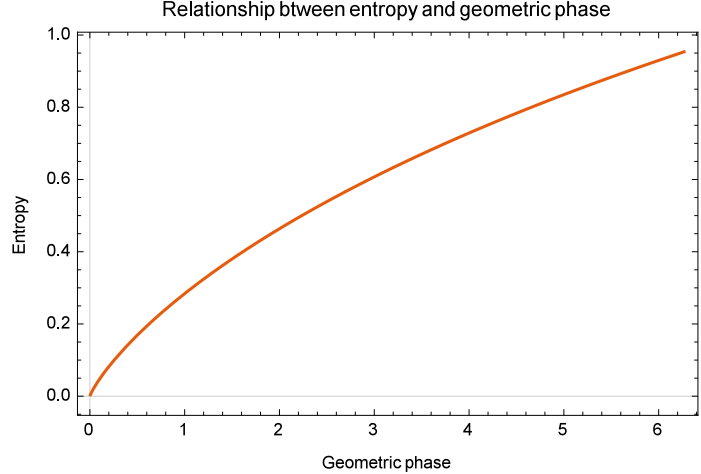}

\caption{\label{fig:entanglement}$\gamma_{c}$ is set to vary form $0$ to
$2\pi$.}
\end{figure}

Moreover, the cyclic goemtric phase $\gamma_{c}$ is related to the
Von Neumann entropy which can meature the entanglement between the
two modes in the squeezed state. In order to establish the relationship.
Let's calculate the entropy first. By use of Eq. \eqref{eq:FullDecompostion},
\[
S(r,\varphi-\Omega t)|0\rangle=\frac{1}{\cosh r}\sum_{n=0}^{\infty}(-e^{2i(\varphi-\Omega t)}\tanh r)^{n}|n\rangle_{+}|n\rangle_{-}.
\]
Fortunetly, it is already in the form of Schmidt decompostion. By
a brute force calulation, the entropy reads
\begin{equation}
E=\cosh^{2}r\ln(\cosh^{2}r)-\sinh^{2}r\ln\sinh^{2}r,\label{eq:Entropy}
\end{equation}
which is identical to the result in Ref. \cite{enk1999discrete}.
Finally, substituting Eq. \eqref{eq:CyclicGeometricPhase} into Eq.
\eqref{eq:Entropy}, we obtain
\[
E=(1+\frac{\gamma_{c}}{4\pi})\ln(1+\frac{\gamma_{c}}{4\pi})-\frac{\gamma_{c}}{4\pi}\ln\frac{\gamma_{c}}{4\pi},
\]
which shows the relationship between the engtanglement and the cyclic
geoemtric phase. And the corresponding graph is in Fig. \eqref{fig:entanglement}.

\section{Conclusions and Acknowledgements }

\label{sec:discussion}

In this article, the geometric phase factor for two-mode squeezed
state is evaluated explicitly. The total phase factor \eqref{eq:TotalPhaseFactorOriginal}
is turned to be an elegant outcome, which is just one term of the
product of the initial squeezed operator and final squeezed operator
\eqref{eq:ProductDecomposition}. When this system undergoes cyclic
evolutions, the corresponding geometric phase is obtained, which is
just the sum of the counterparts of two isolated one-mode squeezed
state. Furthermore, the relationship between the cyclic geomtric phase
and entanglement of two-mode squeezed state is established.

D.B.Y. is supported by NSF of China under Grant No. 11447196. And
J.X.H. is supported by the NSF of China under Grant 11304037, the
NSF of Jiangsu Province, China under Grant BK20130604, as well as
the Ph.D. Programs Foundation of Ministry of Education of China under
Grant 20130092120041.

\bibliographystyle{plain}

\end{document}